\documentclass{ws-procs9x6}
\usepackage{lineno}

\newcommand{\tone}   {\tilde{t}^{}_{1}}

\newcommand{\ttwo}   {\tilde{t}^{}_{2}}

\newcommand{\neut}  {\tilde{\chi}^{0}_{1}} 
\newcommand{\neuttwo}  {\tilde{\chi}^{0}_{2}} 
\newcommand{\charg}  {\tilde{\chi}^{\pm}_{1}}

\newcommand{\bone}   {\tilde{b}^{}_{1}}

\newcommand{\met}{\ensuremath{E_{\text{T}}^{\text{miss}}}}

\begin{document}

\title{Third generation superpartners:\\Results from ATLAS and CMS}

\author{J. Poveda$^*$ on behalf of the ATLAS and CMS Collaborations}

\address{Department of Physics, Indiana University,\\
Bloomington, IN 47405-7105, United States\\
$^*$E-mail: Ximo.Poveda@cern.ch}

\begin{abstract}
These proceedings discuss results from recent searches for third generation SUSY particles 
by the ATLAS and CMS experiments at the LHC. 
Analyses performed with 8 TeV data probing direct pair production 
of bottom and top squarks are presented.
\end{abstract}

\keywords{LHC, ATLAS, CMS, SUSY, stop, sbottom.}

\bodymatter

\section{Introduction}\label{aba:sec1}

After the discovery of a Higgs boson~\cite{Aad:2012tfa,Chatrchyan:2012ufa} in 2012, all the particles included in the Standard Model (SM)
have been observed experimentally. Despite the many successes of this theory in the last few decades, it has several limitations 
from the experimental and theoretical perspectives. For instance, 
the quadratically divergent loop corrections to the Higgs mass would imply a high level of fine-tuning.  
In addition, the SM in its current formulation cannot explain the abundance of dark matter in the universe and
does not provide unification of forces at a high energy scale.

Weak scale supersymmetry 
(SUSY) 
is one of the most studied extensions of the SM and, among other features, can solve all the problems mentioned above. 
SUSY postulates the existence of ``superpartners'' for all the existing SM particles with spin differing by half a unit. 
In $R$-parity conserving SUSY models, the lightest supersymmetric particle (LSP), typically the lightest neutralino ($\neut$), is stable and therefore a good dark matter candidate.

The experimental search for the superpartners of the SM bottom and top quarks, 
known as bottom and top squarks or simply as sbottom ($\bone$) and stop ($\tone$), is one of the most active areas in the search for SUSY. 
These particles can be relatively light, and therefore 
Naturalness arguments, introduced to avoid fine tuning in the theory, predict the sbottom and stop masses to be a few hundred GeV, 
which would therefore be produced with sizeable cross sections at colliders, and produce heavy quarks ($t$ and $b$) in their decays. 


\section{Search for third generation SUSY particles in ATLAS and CMS}\label{aba:sec2}

During 2012, the ATLAS~\cite{Aad:2008zzm} and CMS~\cite{Chatrchyan:2008aa} experiments at the LHC~\cite{Evans:2008zzb} acquired more than 20~fb$^{-1}$ of integrated luminosity from proton-proton collisions at $\sqrt{s}=8$~TeV.

These proceedings will cover recent searches for the direct production 
of stop and sbottom, most of them performed using the full data set recorded in 2012.
In these analyses, the basic elements used to distinguish between stop or sbottom signals and the SM backgrounds are the presence of
leptons (produced in chargino, $\charg$, or top decays), 
jets, $b$-jets (produced in the decays of $\bone$ and top) and missing transverse energy ($\met$) from the escaping LSPs and neutrinos. 
Discriminants combining several of these basic objects are also commonly used, such as the transverse mass of a $W$ candidate, defined as $m_{\rm T}^2=2\cdot p_T^{\ell}\cdot\met\cdot(1 - \cos(\Delta\phi(\ell,\met))$. Some analyses also use more complicated methods, 
such as multivariate techniques or multi-dimensional fits.

Most of the interpretations included in these proceedings are obtained in simplified models~\cite{Alves:2011wf}, 
where only a few particles and decays are considered, assuming that the rest of the spectrum is much heavier and decoupled from the process of interest.

\section{Results on sbottom searches}\label{aba:sec3}

The main sbottom decay modes include the decay to either the lightest or the second lightest neutralino ($\neuttwo$) plus a $b$ quark ($\bone\to b\neut, b\neuttwo$), or to charginos via $\bone\to t\tilde{\chi}_1^{\pm}$. 

\begin{figure}[htb]
\begin{center}
\psfig{file=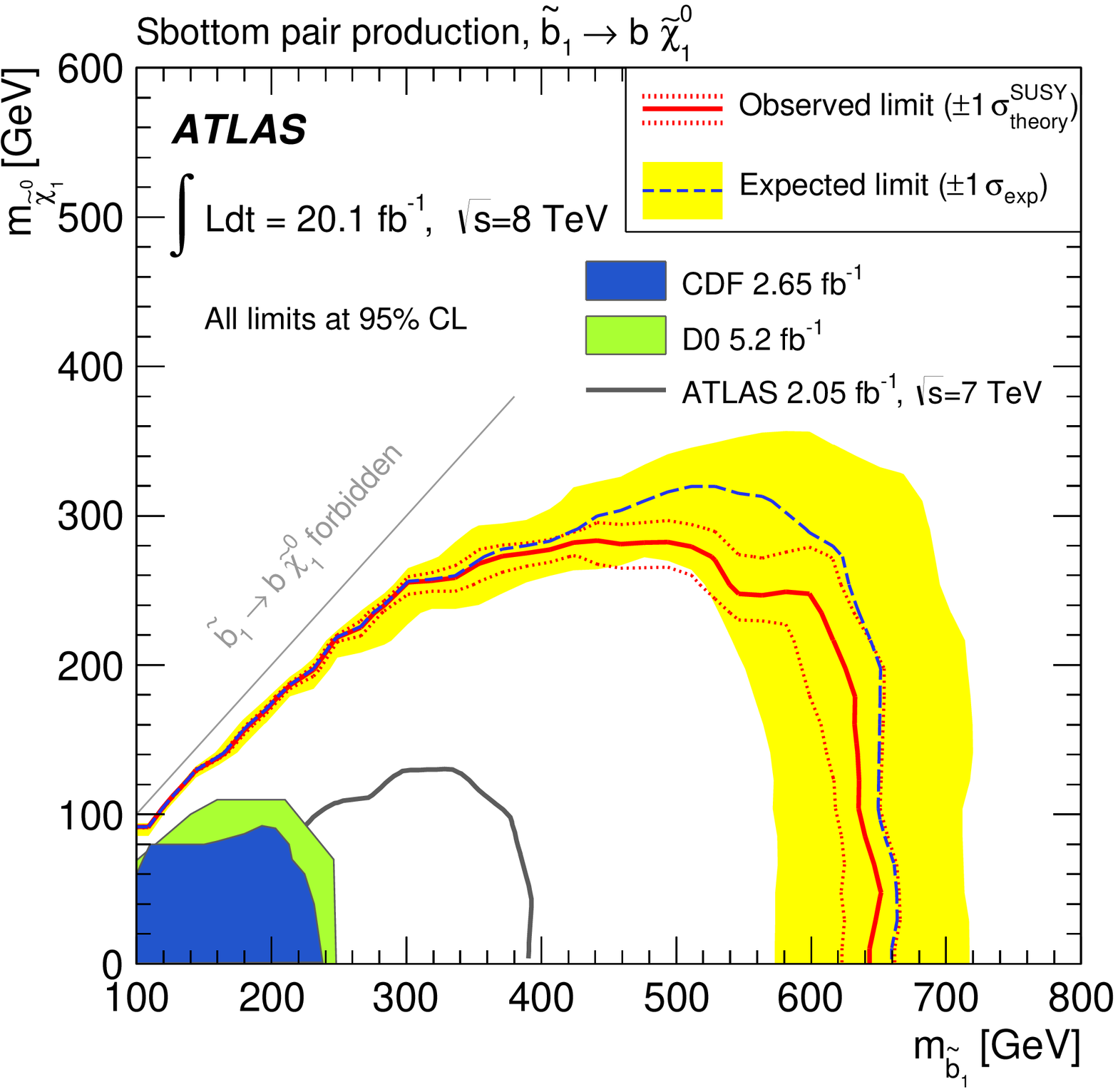,width=2.2in}
\psfig{file=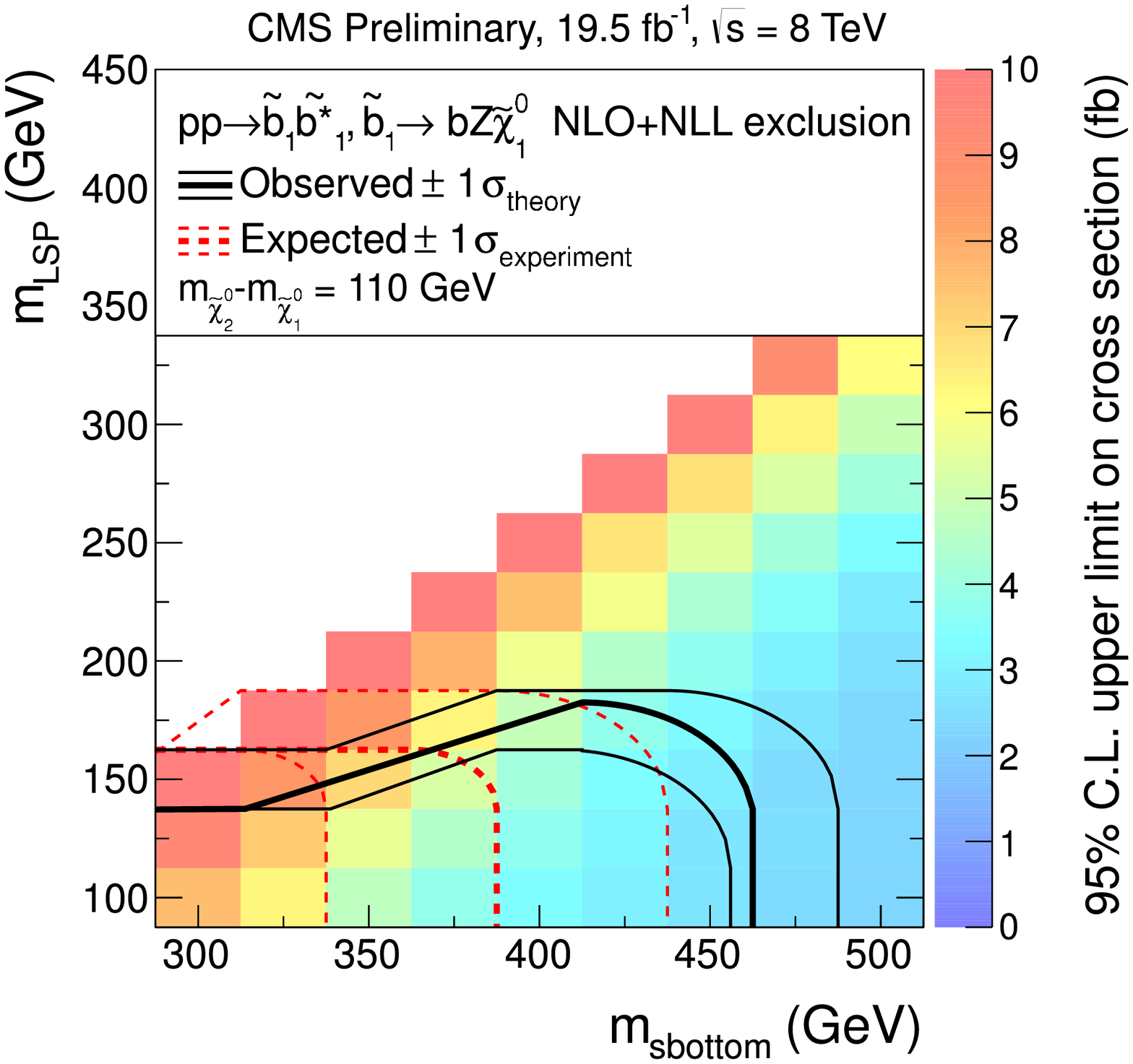,width=2.2in}
\end{center}
\caption{Expected and observed exclusion limits at 95\% CL in the ($m_{\bone}$, $m_{\neut}$) mass plane for simplified models with sbottom pair production. On the left~\cite{Aad:2013ija}, the sbottom decays via $\bone\to b\neut$ and, on the right~\cite{sbotC}, via $\bone\to b\neuttwo$ with $\neuttwo\to Z\neut$.}
\label{aba:fig1}
\end{figure}

In scenarios where the sbottom is produced in pairs and  decays predominantly via $\bone\to b\neut$, 
signal events have a $b\bar{b}\neut\neut$ final state. 
This is explored with analyses requiring a lepton veto, two $b$-jets and $\met$, using event kinematic variables to distinguish 
the sbottom signal from the $b\bar{b}$ background~\cite{Aad:2013ija,Chatrchyan:2013lya}. Figure~\ref{aba:fig1} (left) 
shows an example of the exclusion limits obtained by ATLAS 
for this scenario, reaching sbottom masses of approximately 650~GeV for a massless LSP.

For model characterized by the $\bone\to b\neuttwo$ decay, $Z$ or $h$ bosons are produced via $\neuttwo\to h\neut$ 
or $\neuttwo\to Z\neut$, and the additional 
leptons and $b$-jets in the $h\to b\bar{b}$ or $Z\to\ell\ell$ decays are used to discriminate signal from background~\cite{sbotA,sbotC}. 
Figure~\ref{aba:fig1} (right) shows example exclusion limits obtained by CMS considering a simplified model defined by the $\bone\to b\neuttwo$
with $\neuttwo\to Z\neut$ decay chain, with sensitivity to sbottom masses above 450~GeV.

\begin{figure}[htb]
\begin{center}
\psfig{file=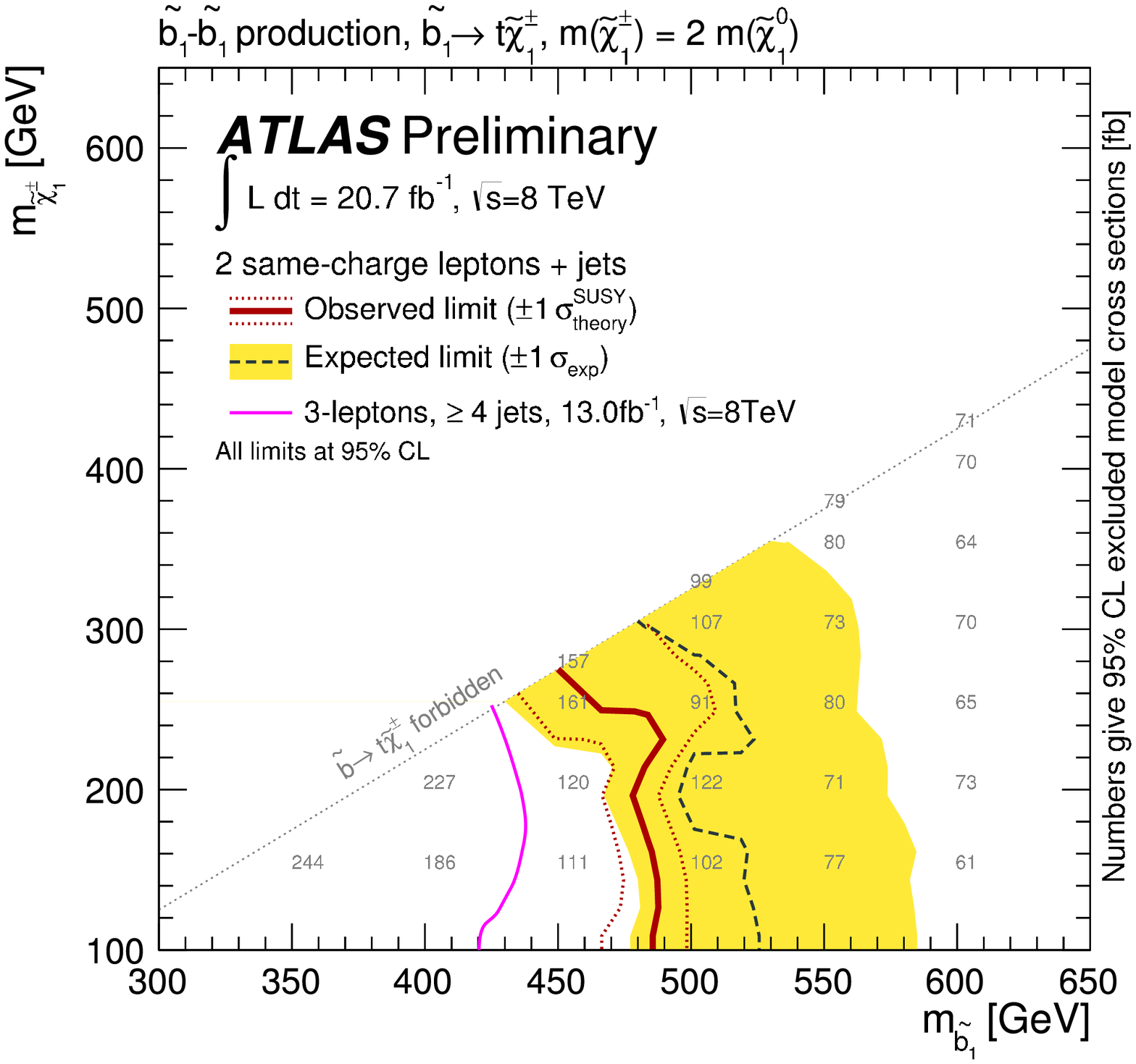,width=2.2in}
\psfig{file=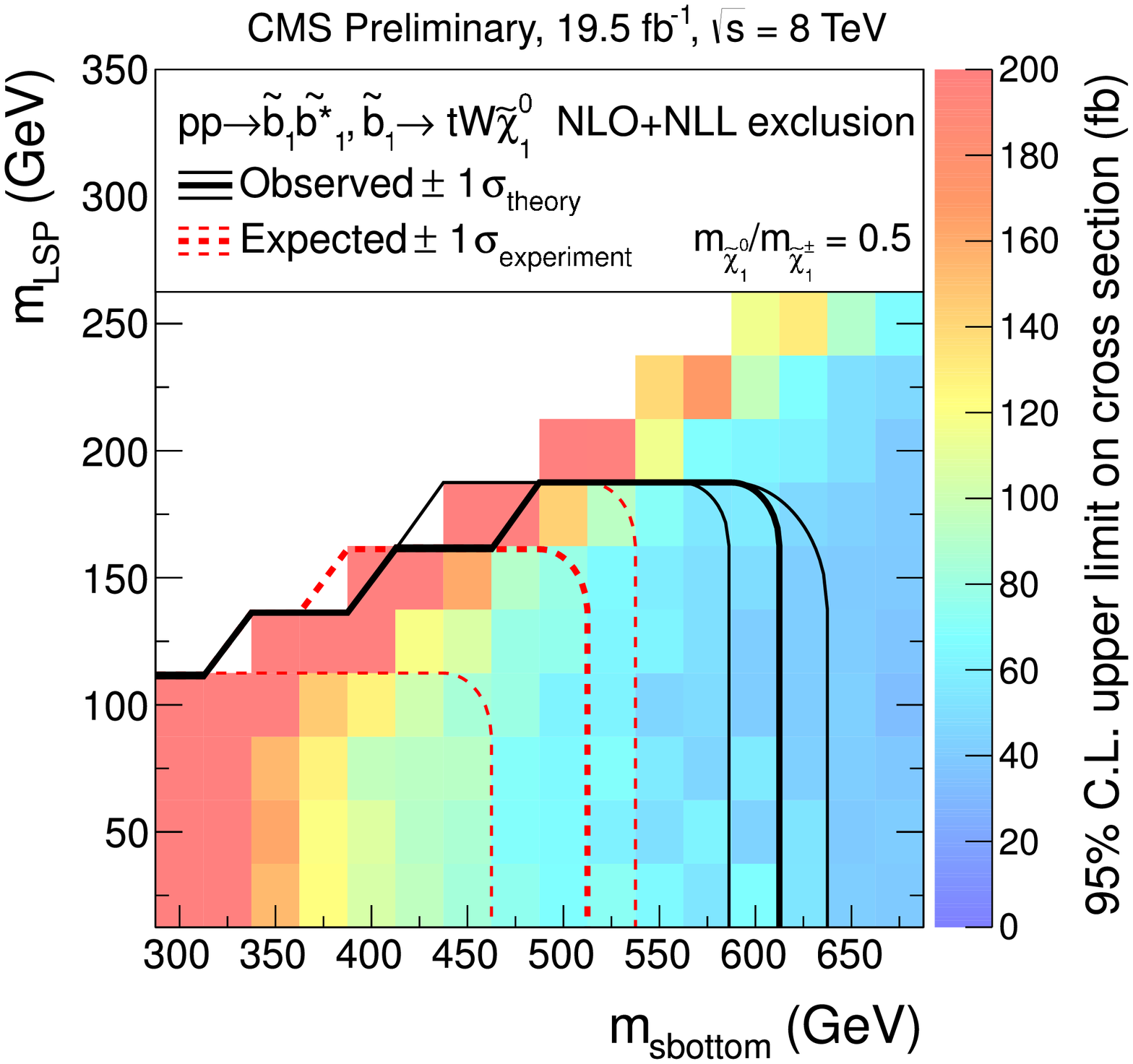,width=2.2in}
\end{center}
\caption{Expected and observed exclusion limits at 95\% CL in the ($m_{\bone}$, $m_{\neut}$) mass plane for simplified models with sbottom pair production and $\bone\to t\tilde{\chi}_1^{\pm}$ decays obtained with dilepton same-sign~\cite{SS_A} (left) and three-lepton~\cite{sbotC} (right) analyses.}
\label{aba:fig2}
\end{figure}

If the sbottom decays predominantly via $\bone\to t\tilde{\chi}_1^{\pm}$, the final state includes additional leptons from the top 
and chargino decays ($\tilde{\chi}_1^{\pm}\to W^{\pm}\neut$). This signal scenario is explored in ATLAS and CMS with analyses requiring two same-sign leptons~\cite{SS_A,SS_C} 
or three leptons~\cite{3L_1,sbotC} in the final state. 
Figure~\ref{aba:fig2} shows the interpretations obtained from some of these analyses for the cases where $m_{\tilde{\chi}_1^{\pm}} = 2 m_{\neut}$, 
with sensitivities for sbottom masses in the range of 500-600~GeV. 

\section{Results on stop searches}\label{aba:sec4}

The decay of the top squark depends on the mass splitting between the stop and its possible decay products,
leading to very different topologies depending on the mass spectrum. 

For a heavy stop, the dominant decays would be $\tone\to t\neut$ (kinematically allowed if $m_{\tone}>m_t+m_{\neut}$) and
$\tone \to b\charg$ (allowed if $m_{\tone}>m_{\charg}+m_b$). 
However, if the $\tone\to t\neut$ decay is kinematically forbidden ($m_{\tone} < m_t +m_{\neut}$), 
the stop would have a three-body decay ($\tone\to bW\neut$) via an off-shell top.
For a lighter stop, when the decays above are forbidden, 
four-body decays ($\tone\to bW^{(*)} \neut$) via off-shell $t$ and $W$ would occur
if $m_{\tone}-m_{\neut}<m_W +m_b$, or the stop would decay to a charm quark via $\tone\to c\neut$ if $m_{\tone}>m_{\neut}+m_c$.

The searches for a heavy stop decaying predominantly via $\tone\to t\neut$ or $\tone \to b\charg$ are designed 
based on the decay of the $W$ boson in the top or chargino decay modes, 
leading to topologies with zero, one or two leptons. 
The semi-leptonic analyses make use of boosted-decision trees~\cite{stop1L_C} at CMS or 
shape fits in the $\met$ and $m_{\rm T}$ variables~\cite{stop1L_A} at ATLAS to distinguish 
a potential stop signal from the $t\bar{t}$ background.
Similarly, analyses probing the fully leptonic final state by ATLAS~\cite{stop2L_1,stop2L_2} make use of the $m_{\rm{T2}}$ variable~\cite{Burns:2008va,Cho:2007dh} as discriminator between signal and background.
Searches for stop are also performed in the fully hadronic channel by both experiments~\cite{stop0L_A,stop0L_C}, showing a very good sensitivity at high stop masses.

A dedicated analysis is also performed by ATLAS aiming at the $\tone\to c\neut$ decay~\cite{stopC} based on charm-tagging techniques and, 
for the case of charm jets too soft to be detected, on a monojet approach relying on an initial state radiation high-$p_{\rm T}$ jet.

\begin{figure}[htb]
\begin{center}
\psfig{file=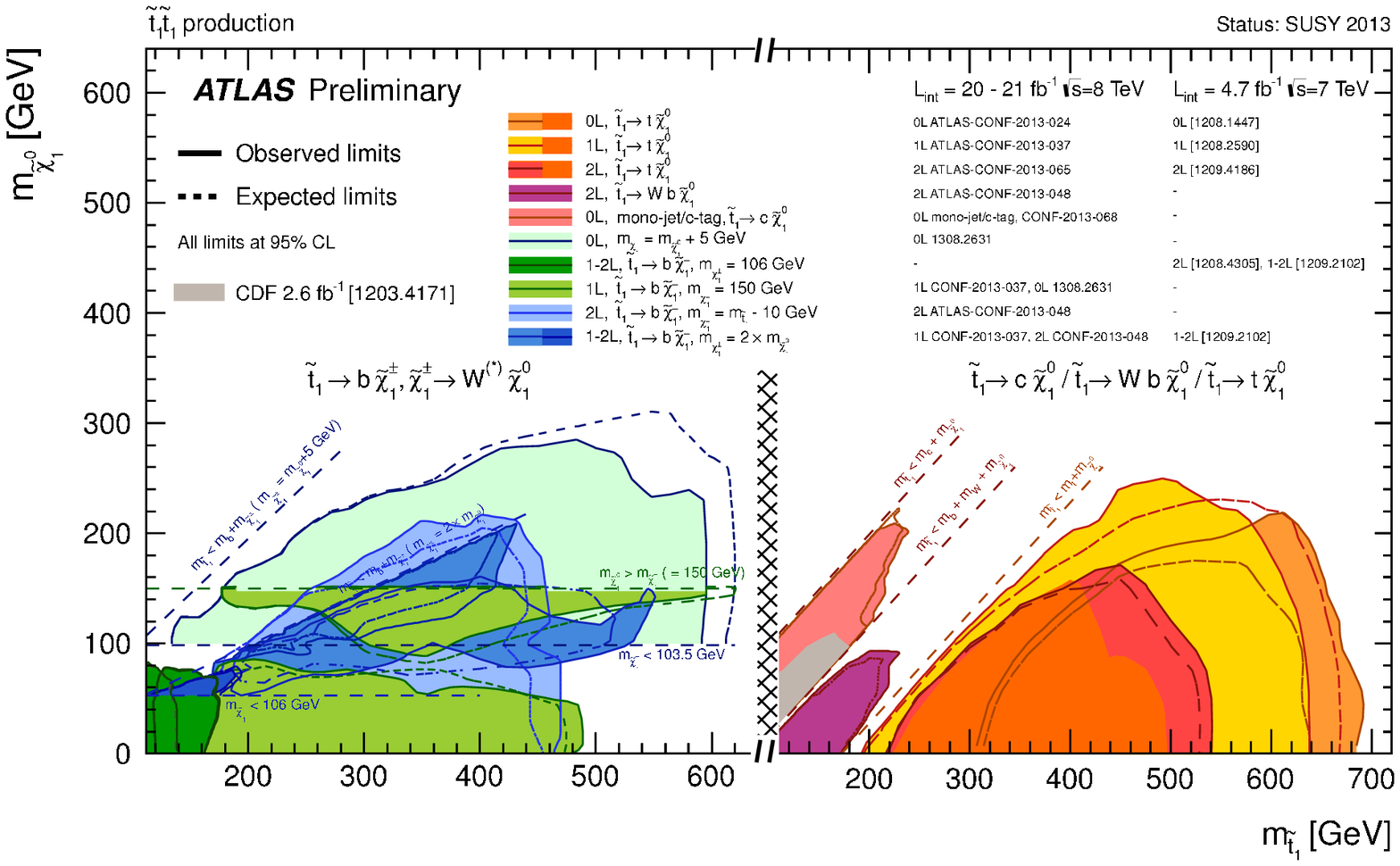,width=4.5in}
\psfig{file=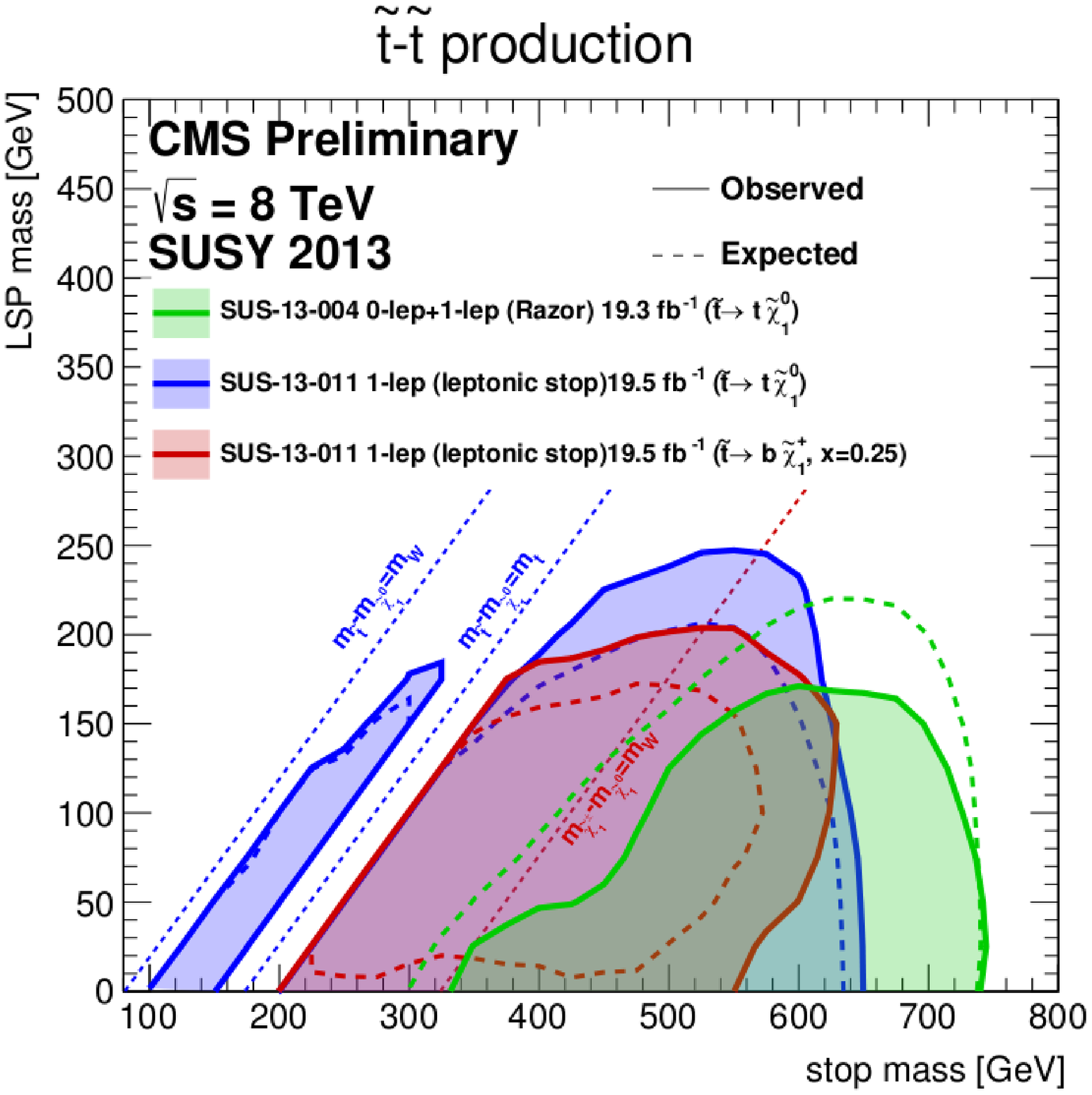,width=2.7in}
\end{center}
\caption{
Summary of the expected and observed exclusion limits at 95\% CL in the ($m_{\tone}$, $m_{\neut}$) from the ATLAS~\cite{ATLAS_plot} (top) and CMS~\cite{CMS_plot} (bottom) searches for stop pair production.}
\label{aba:fig3}
\end{figure}

Figure~\ref{aba:fig3} summarizes the exclusion limits obtained by ATLAS and CMS as a function of the stop and neutralino masses 
for simplified models with different stop decays. Sensitivity up to stop masses of around 700~GeV are obtained for a massless 
neutralino, 
while for massive neutralinos of 250-300~GeV, stop sensitivity falls to 450-500~GeV.

\begin{figure}[htb]
\begin{center}
\psfig{file=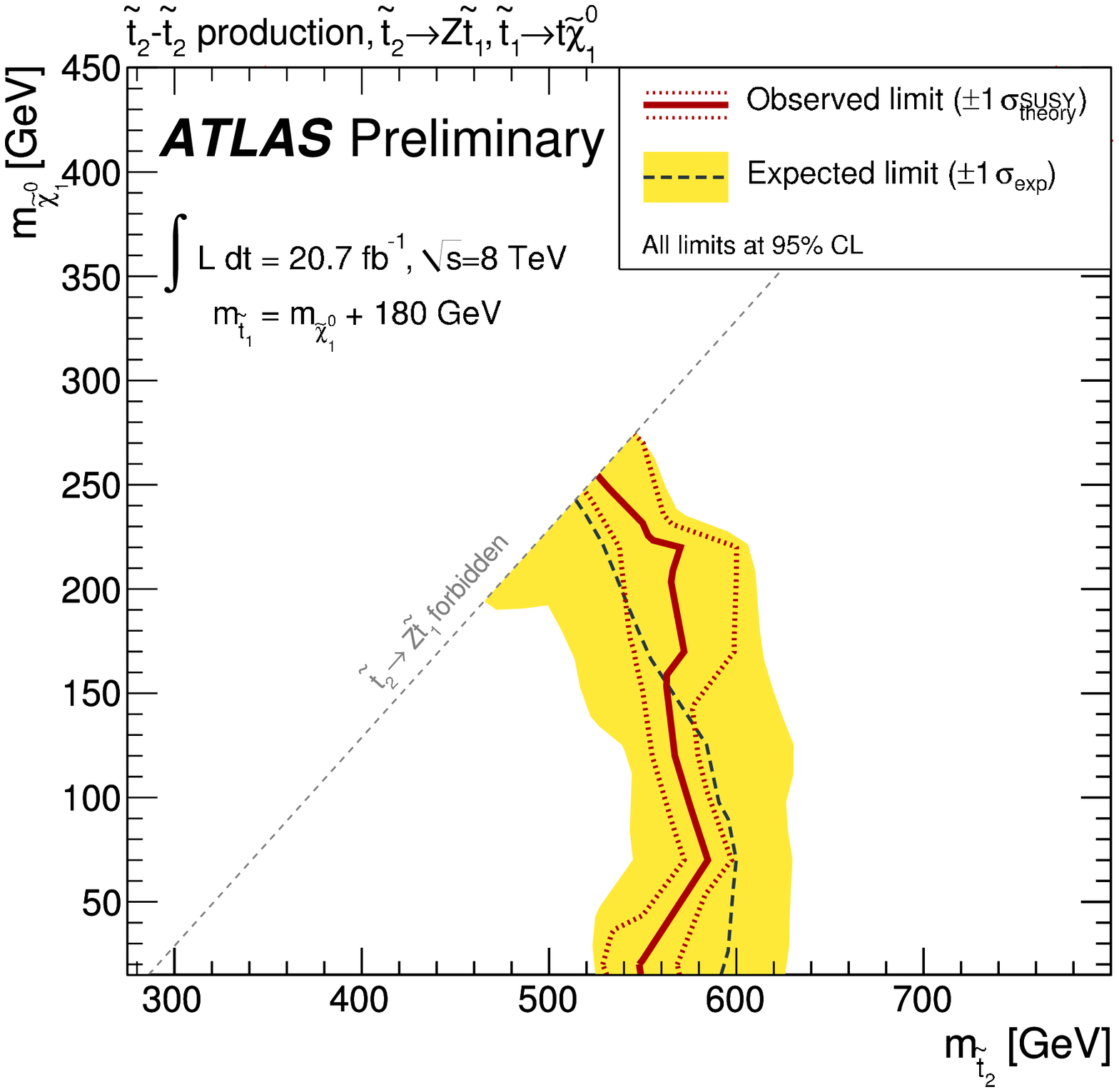,width=2.05in}
\psfig{file=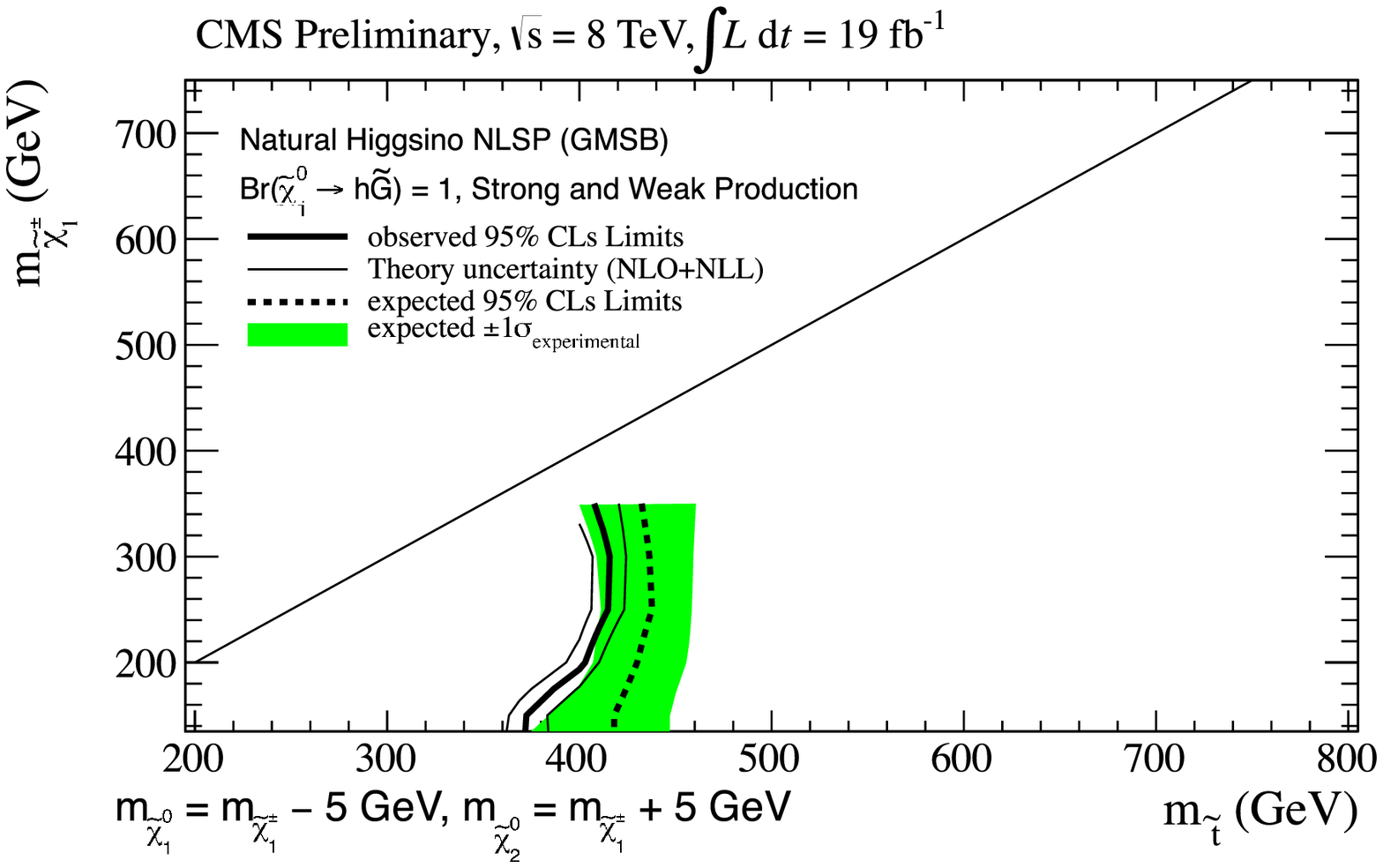,width=2.4in}
\end{center}
\caption{On the left~\cite{stopZ}, expected and observed exclusion limits at 95\% CL obtained in the ($m_{\ttwo}$, $m_{\neut}$) mass plane 
for simplified models with $\ttwo$ pair production and $\ttwo\to Z\tone$, $\tone\to t\neut$ decays for $m_{\tone}\approx m_t + m_{\neut}$. 
On the right~\cite{stopGM_C}, expected and observed exclusion limits at 95\% CL considering a ``natural'' SUSY scenario with gauge-mediated breaking
as a function of $m_{\tone}$ and $m_{\charg}$.}
\label{aba:fig4}
\end{figure}

Despite the good coverage of the stop mass parameter space in Figure~\ref{aba:fig3}, some experimentally challenging regions remain unconstrained.
For example, no exclusion is achieved in the $m_{\tone}\approx m_t + m_{\neut}$ region, where the stop events are kinematically very similar to SM $t\bar{t}$ production. However, this region is explored by ATLAS considering the production of the heavy stop state ($\ttwo$) decaying via $\ttwo\to Z\tone$~\cite{stopZ}. As shown in Figure~\ref{aba:fig4}, exclusion limits on $\ttwo$ masses of approximately 550~GeV are obtained in this very difficult ($m_{\tone}$, $m_{\neut}$) configuration.

Other models also feature long decay chains involving stops, such as gauge-mediated SUSY breaking 
scenarios where the graviton ($\tilde{G}$) is the LSP and the lightest neutralino decays via $\neut\to Z\tilde{G}$ or $\neut\to h\tilde{G}$. 
Analyses exploiting the $Z\to\ell\ell$ or $h\to\gamma\gamma$ decays in this kind of model~\cite{stopZ,stopGM_C,stopGM_C2} 
lead to sensitivities at stop masses of a few hundred GeV, as illustrated in Figure~\ref{aba:fig4} (right) by the CMS analysis 
considering $\neut\to h\tilde{G}$ with $h\to\gamma\gamma$.

\section{Summary and conclusions}\label{aba:sec5}

The search for SUSY is a major part of the LHC physics programme. In particular, naturalness arguments
favour the existence of top and bottom squarks with masses of several hundred GeV which can be observed at the LHC energies.

A wide program of searches for third generation squarks has been carried out in ATLAS and CMS with the LHC data acquired during 2010-2012. 
Dedicated analyses optimized for different stop and sbottom decay modes and final state topologies have been conducted.
In some cases, complex analysis techniques (boosted-decision trees, multi-dimensional fits, $c$-jet tagging, etc.) are used to discriminate 
signals from the background.

No excess has been observed in data above the Standard Model predictions and the current exclusion limits reach stop and sbottom 
masses at the level of 600-700~GeV in models with one step decays such as $\bone\to b\neut$ or $\tone\to t\neut$. 
Very recent results also address the case of long stop and sbottom decay chains involving $Z$ and Higgs bosons.
 
\bibliographystyle{ws-procs9x6}
\bibliography{references}

\end{document}